\documentstyle[12pt,a4]{article}

\newcommand{\nit}{\noindent}
\newcommand{\nl}{\newline}
\newcommand{\np}{\newpage}
\newcommand{\be}{\begin{equation}}
\newcommand{\ee}{\end{equation}}
\newcommand{\ba}{\begin{array}}
\newcommand{\ea}{\end{array}}
\newcommand{\dsp}{\displaystyle}
\newcommand{\ct}{\cite}
\newcommand{\bit}{\bibitem}

\newcommand{\th}{\theta}
\newcommand{\kg}{\kappa}
\newcommand{\lb}{\lambda}
\newcommand{\sg}{\sigma}
\newcommand{\vf}{\varphi}

\newcommand{\Gam}{\Gamma}
\newcommand{\Lb}{\Lambda}
\newcommand{\lh}{\left(}
\newcommand{\rh}{\right)}
\newcommand{\pl}{\partial}
\newcommand{\cL}{{\cal L}}

\begin{document}

\pagestyle{empty}

 \begin{flushright}
 NIKHEF/95-008
 \end{flushright}

\begin{center}
{\Large {\bf DOMAIN BOUNDARIES, GOLDSTONE BOSONS}} \\
\vspace{3ex}

{\Large {\bf AND GRAVITATIONAL WAVES}} \\
\vspace{5ex}

{\large J.W.\ van Holten}\\
\vspace{3ex}

{\large NIKHEF, P.O.\ Box 41882}\\
\vspace{3ex}

{\large Amsterdam NL}
\vspace{5ex}

{\today}
\end{center}
\vspace{15ex}

\nit
{\small
{\bf Abstract}\nl
The dynamical behaviour of domain boundaries between different realizations
of the vacuum of scalar fields with spontaneously broken phases is
investigated. They correspond to zero-modes of the Goldstone fields, moving
with the speed of light, and turn out to be accompanied by strongly
oscillating gravitational fields. In certain space-time topologies
this leads to a quantization condition for the symmetry breaking scale in
terms of the Planck mass. }

\np

\pagestyle{plain}
\pagenumbering{arabic}

\nit
{\bf 1.\ Spontaneous symmetry breaking and domain boundaries}
\vspace{2ex}

\nit
Spontaneous violation of rigid continuous symmetries in a field theory
implies the existence of massless scalar excitations \ct{JG,NJL}. A
simple example is provided by the $U(1)$ invariant scalar model

\be
\cL\, =\, - \pl_{\mu} \vf^* \pl^{\mu} \vf\, -\,
        \frac{\lb}{4}\, \lh \vf^* \vf - \mu^2 \rh^2.
\label{1}
\ee

\nit
If $\mu^2 > 0$ we can parametrize the classical minima of the potential as

\be
\vf\, =\, \mu\, e^{i \th}.
\label{2}
\ee

\nit
The angle $\th$ is an unobservable parameter. However, if $\th$ differs
between various locations in space-time, the local variations of $\th$
are observable. Indeed, as is well-known allowing the angle to become a
space-time dependent field $\th (x) = \sg (x)/\mu\sqrt{2}$, the regular
excitations of this field correspond to free, massless scalar bosons:

\be
\cL\, =\, - \frac{1}{2}\, \pl_{\mu} \sg \pl^{\mu} \sg,
\label{3}
\ee

\nit
with the associated field equation

\be
\Box\, \sg\, =\, 0.
\label{4}
\ee

\nit
Even if the scalar field is frozen in a classical minimum (\ref{2}) over
finite, extended regions of space-time, the dynamics of the fields can
still cause the value of $\th$ to differ in different regions. Such a
situation is conceivable in cosmological models of the large-scale
structure of the universe \ct{AV}.

The boundaries between such regions correspond to zero-modes of the
field equation (\ref{4}):

\be
\sg(x)\, =\, \sg_0\, +\, \mu \sqrt{2}\, k \cdot x, \hspace{3em}
             k_{\mu}^2\, =\, 0.
\label{5}
\ee

\nit
In the simplest case we have a flat boundary of finite thickness $L$,
for example in the $(y,z)$-plane, with on one side $(x = + \infty)$ a
classical vacuum field configuration with arbitrary constant angle $\th_0$,
and on the other $(x = - \infty)$ a similar configuration with constant
angle $\th_1$. Then the solution in the boundary region is

\be
\sg(x)\, =\, \sg_0\, +\, \mu \sqrt{2}\, k \lh x - t \rh, \hspace{3em}
             - L \leq \lh x - t \rh \leq 0.
\label{6}
\ee

\nit
Such a boundary corresponds to a  right-moving collective excitation, moving
at the speed of light\footnote{In natural units $c = 1$. Of course there is
an equally valid solution moving to the left.}. The value of the wave number
$k$ is

\be
k\, =\, \frac{\th_1 - \th_0}{L}\, =\, \frac{\sg_1 - \sg_0}{\mu L \sqrt{2}}\, ,
\label{7}
\ee

\nit
and the energy density in the wave ${\cal E} = 2\mu^2 k^2$.

The Noether current for the $U(1)$ symmetry of $\cL$ is

\be
j_{\mu}\, =\, - \frac{i}{2}\, \vf^* \stackrel{\leftrightarrow}{\pl}_{\mu} \vf\,
          =\, \frac{\pl_{\mu} \sg}{ \mu \sqrt{2} }\, .
\label{8}
\ee

\nit
It is conserved by the equation of motion (\ref{4}).
\vspace{3ex}

\nit
{\bf 2.\ Coupling to gravity}
\vspace{2ex}

\nit
In cosmological applications, the Goldstone fields have to be coupled to
gravity. The action becomes

\be
S\, =\, \int d^4x\, \sqrt{-g}\, \lh \frac{-1}{16 \pi G}\, R\, -\,
        \frac{1}{2}\, g^{\mu\nu} \pl_{\mu} \sg \pl_{\nu} \sg \rh.
\label{9}
\ee

\nit
It turns out, that the solutions (\ref{5}) remain solutions also in the
presence of gravity. However, the finite energy density ${\cal E}$ of the
boundary moving at the speed of light now becomes a source of gravitational
waves. These waves correspond to oscillations of the gravitational field
which travel together with the boundary at the speed of light and remain
associated with it. In fact, the solution we present below, in which an
oscillating gravitational field is associated with a constant Goldstone
current, has certain similarities with the Josephson effect in the
theory of superconductivity.

For the system (\ref{9}) the Einstein equations take the form

\be
R_{\mu\nu}\, =\, - 8 \pi G \pl_{\mu} \sg \pl_{\nu} \sg,
\label{10}
\ee

\nit
whilst the scalar field equation becomes

\be
\Box^{cov}\, \sg\, =\, \frac{1}{\sqrt{-g}}\, \pl_{\mu}
                       \sqrt{-g} g^{\mu\nu} \pl_{\nu}\, \sg\, =\, 0.
\label{11}
\ee

\nit
For the solutions (\ref{6}) to remain valid, we make the following {\em
Ansatz} for the metric

\be
ds^2\, =\, - dt^2\, +\, dx^2\, +\, f^2 dy^2\, + g^2 dz^2,
\label{12}
\ee

\nit
with the additional restriction that $f = f(x -t)$ and $g = g(x-t)$.
At this point the restriction is actually stronger than necessary,
but this form is also required by the Einstein equations; therefore
we impose it already here.

With the metric (\ref{12}) the Einstein equations
reduce to a single additional constraint

\be
\frac{f^{\prime\prime}}{f}\, +\, \frac{g^{\prime\prime}}{g}\, =\,
  - 16 \pi G \mu^2 k^2.
\label{13}
\ee

\nit
Up to trivial relabeling of co-ordinates, this equation admits
two families of solutions. If we take $f = g = 1$ for $x - t \geq 0$,
then inside the boundary region $-L \leq (x - t) \leq 0$ the first class
of solutions is

\be
f\, =\, \cos \kg \lh x - t \rh, \hspace{3em}
g\, =\, \cos \lb \lh x - t \rh.
\label{14}
\ee

\nit
The parameters $(\kg, \lb)$ are related by

\be
\kg^2\, +\, \lb^2\, =\, 16 \pi G \mu^2 k^2.
\label{15}
\ee

\nit
These solutions will be called elliptic. Note that the solutions are
continuous and differentiable at $x - t = 0$. They are generalizations of the
plane fronted waves discussed for example in \ct{BPR,EK}. The solutions can
also remain continuous  and differentiable at $x - t = -L$, provided for
$x - t \leq -L$ we take

\be
\ba{lll}
f(x - t) & = & \cos \kg L\, +\, \kg \lh L + x - t \rh\, \sin \kg L, \\
 & & \\
g(x - t) & = & \cos \lb L\, +\, \lb \lh L + x - t \rh\, \sin \lb L. \\
\ea
\label{16}
\ee

\nit
Such solutions linear in $(x-t)$ correspond to flat space-time,
because all components of the Riemann tensor vanish.

Clearly, for $\th_1 \rightarrow \th_0$ one has $k \rightarrow 0$ and
as a consequence $(\kg, \lb) \rightarrow 0$. Then space-time becomes
flat everywhere: $(f,g) \rightarrow 1$. This is not true in the other
class of solutions, which are of the type

\be
f\, =\, \cos \kg \lh x - t \rh, \hspace{3em}
g\, =\, \cosh \lb \lh x - t \rh,
\label{17}
\ee

\nit
with the additional constraint

\be
\kg^2\, - \lb^2\, =\, 16 \pi G \mu^2 k^2.
\label{18}
\ee

\nit
These solutions we call hyperbolic. In this case the flat space
behind the wave is parametrized as

\be
\ba{lll}
f(x - t) & = & \cos \kg L\, +\, \kg \lh L + x - t \rh\, \sin \kg L, \\
 & & \\
g(x - t) & = & \cosh \lb L\, -\, \lb \lh L + x - t \rh\, \sinh \lb L. \\
\ea
\label{19}
\ee

\nit
In the limit $k \rightarrow 0$ a plane fronted purely gravitational wave
remains, with $\kg = \lb$.
\np

\nit
{\bf 3.\ Compactification and the quantization of the symmetry breaking scale}
\vspace{2ex}

\nit
An interesting consequence of the previous results is, that if space-time
can be compactified on a cylinder such that the Minkowski spaces at $x =
\pm \infty$ can be identified, this requires a quantization condition

\be
\kg\, =\, \frac{2 \pi n}{L}, \hspace{3em}
\lb\, =\, \frac{2 \pi m}{L},
\label{20}
\ee

\nit
for the elliptic solutions, and

\be
\kg\, =\, \frac{2 \pi n}{L}, \hspace{3em}
\lb\, =\, 0,
\label{21}
\ee

\nit
for the hyperbolic solutions. At the same time, the scalar phase angles
at $x = \pm \infty$ must be equal mod $2 \pi$ as well; equivalently:

\be
k\, =\, \frac{\th_1 - \th_0}{L}\, =\, \frac{2 \pi l}{L}.
\label{22}
\ee

\nit
As a result we get for the elliptic case

\be
16 \pi G \mu^2\, =\, \frac{n^2 + m^2}{l^2}.
\label{23}
\ee

\nit
By taking $m = 0$ this formula holds for the hyperbolic solutions as well.
Thus a relation between Newton's constant and the symmetry breaking scale
is obtained in a natural way. It can also be expressed in terms
of the Planck mass as

\be
\mu^2\, =\, \frac{n^2 + m^2}{l^2}\, \frac{M_{Pl}^2}{16 \pi}.
\label{24}
\ee

\nit
{\bf 4.\ Orbits of test masses}
\vspace{2ex}

\nit
In principle the oscillations of the metric components in equations (\ref{14})
and (\ref{17}) can be observed from the behaviour of test masses in the
laboratory. To show this, we introduce a laboratory co-ordinate
frame $X^{\mu}$ by defining

\be
\ba{ll}
\dsp{ T\, =\, t - \frac{\Lb}{2},} & \dsp{ X\, =\, x - \frac{\Lb}{2}, } \\
  & \\
Y\, =\, f y, & Z\, =\, g z,
\ea
\label{25}
\ee

\nit
where

\be
\Lb\, =\, y^2 f f^{\prime}\, +\, z^2 g g^{\prime}\,
      =\, \frac{f^{\prime}}{f}\, Y^2\, +\, \frac{g^{\prime}}{g}\, Z^2.
\label{26}
\ee

\nit
Moreover, it is useful to introduce light-cone co-ordinates in both frames
by taking

\be
U\, =\, X\, -\, T\, =\, u, \hspace{3em}
V\, =\, X\, +\, T\, =\, v - \Lb.
\label{27}
\ee

\nit
Then the line element (\ref{12}) becomes

\be
\ba{lll}
ds^2 & = & \dsp{ dU dV\, +\, dY^2\, +\, dZ^2\, +\, \lh
           \frac{f^{\prime\prime}}{f}\, Y^2 +
           \frac{g^{\prime\prime}}{g}\, Z^2 \rh dU^2 }
\ea
\label{28}
\ee

\nit
Clearly, in the regions of constant scalar phase: $x - t \geq 0$ or
$ x - t \leq -L$, the co-efficient of $dU^2$ vanishes and we have manifestly
Minkowskian regions of space-time. In the boundary region $-L < x- t < 0$ the
line element reads explicitly

\be
\ba{lll}
ds^2 & = & dU dV\, +\, dY^2\, +\, dZ^2\, -\, \lh \kg^2 Y^2 \pm
           \lb^2 Z^2 \rh dU^2,
\ea
\label{29}
\ee

\nit
with the plus/minus sign corresponding to elliptic and hyperbolic solutions,
respectively. In these co-ordinates the Riemann tensor is constant:

\be
R_{U Y U Y}\, =\, - \kg^2, \hspace{3em}
R_{U Z U Z}\, =\, \mp \lb^2,
\label{29.1}
\ee

\nit
whilst all other independent components vanish. The metric does not oscillate,
but test particles do with respect to the co-ordinate frame. This can be seen
from the explicit solutions of the equations of motion

\be
\frac{d^2 X^{\mu}}{d\tau^2}\, +\, \Gam_{\lb\nu}^{\:\:\:\:\:\mu}\,
\frac{dX^{\lb}}{d\tau}\, \frac{dX^{\nu}}{d\tau}\, =\, 0.
\label{30}
\ee

\nit
Using the line element (\ref{29}) and choosing the unit of co-ordinate time
in the origin $(Y_0 = Z_0 = 0)$ equal to the unit of proper time, the elliptic
solutions read:

\be
\ba{ll}
U = U_0 - \tau, & \dsp{ V = V_0 + \tau - \frac{1}{2}\, \lh
    \kg Y_0^2 \sin 2 \kg \tau + \lb Z_0^2 \sin 2 \lb \tau \rh, }\\
  & \\
Y = Y_0 \cos \kg \tau, & Z\, =\, Z_0 \cos \lb \tau.
\ea
\label{31}
\ee

\nit
For the hyperbolic solutions we replace $\cos \lb \tau$ by
$\cosh \lb \tau$, and $\sin 2 \lb \tau$ by $ - \sinh 2 \lb \tau$.
The oscillatory motion of the test masses with respect to the origin
$Y = Z = 0$ is manifest in this laboratory frame.
\vspace{3ex}

\nit
{\bf 5.\ Discussion}
\vspace{2ex}

\nit
We have shown that plane fronted gravitational waves are present in regions
where
Goldstone bosons have constant gradients. Physically such a situation is
present in boundary regions between domains of constant $\sg$. In these regions
the original scalar field $\phi$ oscillates as well:

\be
\phi(x)\, =\, \mu\, e^{i k \cdot x}.
\label{32}
\ee

\nit
The frequency of these scalar waves is much higher in general than that of the
associated gravitational wave: $k \gg \kg$, unless the vacuum expectation $\mu$
is of the order of the Planck mass.  Therefore the frequency of the
gravitational waves becomes appreciable only for very large values of $k$, or
very steep changes in the value of the phase $\th$. This applies for very thin
boundaries.

Finally we would like to point out that instead of (massless) scalar potentials
with constant gradients, one might also consider massless vector fields with
constant field strength $(E, B)$. The generation of gravitational waves in
constant electromagnetic fields has been discussed elsewhere \ct{JW}.

\end{document}